\documentclass[pre,twocolumn,showpacs,floatfix]{revtex4-1}
\usepackage{epsfig}
\usepackage{color}

\begin{document}

\title{Edwards entropy and compactivity in a model of granular matter.}

\author{Richard K. Bowles}\thanks{Corresponding Author}
\affiliation{Department of Chemistry, University of Saskatchewan,
Saskatoon, Saskatchewan, S7N 5C9, Canada}
\email{richard.bowles@usask.ca}

\author{S. S. Ashwin}
\affiliation{Department of Chemistry, University of Saskatchewan,
Saskatoon, Saskatchewan, S7N 5C9, Canada}

\begin{abstract}
Formulating a statistical mechanics for granular matter remains a significant challenge, in part, due to the difficulty associated with a complete characterization of the systems under study. We present a fully characterized model of a granular material consisting of $N$ two-dimensional, frictionless, hard discs, confined between hard walls, including a complete enumeration of all possible jammed structures. We show the properties of the jammed packings are independent of the distribution of defects within the system and that all the packings are isostatic. This suggests the assumption of equal probability for states of equal volume, which provides one possible way of constructing the equivalent of a microcanonical ensemble, is likely to be vaild for our model. An application of the second law of thermodynamics involving two subsystems in contact shows that the expected spontaneous equilibration of defects between the two is accompanied by an increase in entropy and that the equilibirum, obtained by entropy maximization, is characterized by the equality of compactivities. Finally, we explore the properties of the equivalent to the canonical ensemble for this system.
\end{abstract}


\maketitle

% introduction
\section{Introduction}
Granular materials such as sand are too massive to be influenced by the thermal fluctuations that affect the motion of particles on the atomic scale. If left undisturbed, a sand pile remains in a rigid, solid-like structure, but when poured, sand flows just like a fluid. Foams, made up of a collection of macroscopic bubbles, can support a yield stress like a solid but they flow like a fluid when sheared above a certain threshold rate. In both cases, the particle-particle structure and particle dynamics are similar to that observed in atomic and molecular liquids and glasses, suggesting that many athermal systems may exhibit a wide variety of physical phenomena that were originally thought to occur only in thermal systems~\cite{gen99}. While there have been a number of efforts to capture the connection between the different systems, including the formulation of a phase diagram for jammed and glassy materials~\cite{liu98}, we do not have an underlying and general theory to describe granular matter. The machinery of classical thermodynamics and thermal statistical mechanics, which provides us with the tools for the study of complex behavior in molecular systems, cannot be used to explain the behavior of a granular system because the energy required to move a macroscopic particle far exceeds that available from thermal excitation. Consequently, there is an extensive effort to develop a form of statistical mechanics that can account for the properties of athermal systems~\cite{ed1,ed2,blu06,mak08N,mak08,ash09,hen07}.

% Statistical mechanics.
An appropriate starting point  for a statistical mechanics of granular systems is to develop a probability distribution function that is equivalent to the microcanonical ensemble where all the states of the same energy are assumed to have an equal probability of being sampled. One possibility, suggested by Edwards~\cite{ed1,ed2}, is to assume that all jammed states of equal volume have equal probabilities so that the probability of finding a given configuration $i$ is,
\begin{equation}
P_i=e^{-S/\lambda}\delta(V-W_i)\Theta\mbox{,}
\label{eq:mcp}
\end{equation}
where $V$ is the volume, $W_i$ is the Hamiltonian-like volume function for the configuration, $S$ is the entropy and $\lambda$ is the analogue of the Boltzmann constant. The normalization in Eq.~\ref{eq:mcp} is
\begin{equation}
e^{S/\lambda}=\int\delta (V-W_i) \Theta d\mbox{(all degrees of freedom)}\mbox{,}
\label{eq:s1}
\end{equation}
and
\begin{equation}
\Theta=\left\{\begin{array}{ll}
1& \mbox{if collectively jammed}\\
0&\mbox{otherwise}\end{array}
\right.\mbox{.}
\label{eq:jam}
\end{equation}
Evaluating the integral in Eq.~\ref{eq:s1} and taking the natural log of both sides defines the entropy as,
\begin{equation}
S/\lambda=\ln \Omega(V)\mbox{,}\\
\label{eq:s2}
\end{equation}
where $\Omega(V)$ is the number of collectively jammed configurations~\cite{tor01,don05,don05b} with a given $V$.  $\Omega(V)$ grows exponentially with the number of particles, $N$, so that entropy is extensive and the compactivity, which is equivalent to the temperature in a thermal system, can be defined as~\cite{ed1,ed2},
\begin{equation}
X=\frac{\partial V}{\partial S}\mbox{.}\\
\label{eq:x1}
\end{equation}

% Problems of obtaining distribution
In general, $\Omega(V)$ is not known. One difficulty is that little is known about how particles share volume in jammed materials and while a number of different tessellation methods have been suggested~\cite{blu06, mak08N,mak08,ash09}, the construction of $W$ remains a significant challenge. Furthermore, the global nature of the collective jamming condition~\cite{tor01,don05,don05b} does not allow us to determine which configurations are jammed on the basis of the local geometric properties of the individual particles. The effects of friction and stress in jammed configurations add further complications, leading to the proposal of alternative distribution functions~\cite{hen07}.

A numerical simulation of a slowly, sheared granular material~\cite{mak02} suggests a thermodynamic temperature, related to compactivity, can be extracted from particle mobility measurements, but simulations of frictionless hard discs~\cite{gao06} appear to cast doubt on the assumption that packings of equal volume are equi-probable. Recent experiments~\cite{dan09} on externally-agitated granular systems found that while compactivity did not equilibrate between two subsystems, both subsystems shared a number of reproducible properties, strongly supporting the notion that a thermodynamic approach to granular materials is a real possibility, even though we do not have a complete picture of how this should be developed. 

However, for all the systems studied to-date, the distribution function and a detailed knowledge of the particle packings is lacking. The goal of the present work is to examine the statistical mechanics of a simple model granular system that has the potential to be examined experimentally and for which the full density states can be obtained exactly. We analyze the distribution of jammed packings for a system of highly confined hard discs in Section II and show that the assumption of equal probability for states of equal volume is likely to be valid for this model. In Section III, we examine the predictions of the second law of thermodynamics by considering the equilibrium state obtained by bringing together two isolated systems into the granular equivalent of thermal contact. The canonical ensemble is studied in Section IV while Section V contains our discussion and conclusions.

%id landscape
\section{Packings of highly confined hard discs}
Our model granular material consists of $N$ two-dimensional ($2D$) frictionless, hard discs, with diameter $\sigma$, trapped between two hard walls separated by a distance $H/\sigma<1+\sqrt{3/4}$ and we will ignore the effects of gravity. The density of states for this system was previously described in Ref.~\cite{rkb06} and is included here for the sake of completeness and to highlight the fact that the distribution of packings for this system can be obtained using simple combinatorial arguments. In addition, we also show here that all the packings of the system are isostatic and that packings of the same volume are all structurally equivalent. 

In $2D$, a particle is locally jammed if it has at least three rigid contacts that are not all in the same semicircle. However, while local jamming of all particles is a necessary requirement for a collectively jammed state, it is not sufficient because the concerted motion of a group of particles can cause a packing to collapse. By confining the discs between walls separated by $H/\sigma<1+\sqrt{3/4}$, particles are only able to contact their nearest neighbor on each side and the wall so there are only two  local packing environments. In addition, the confinement prevents collective motions that would allow the locally jammed structures to collapse and the complete distribution of collectively jammed states can be obtained from local geometric considerations alone.

\begin{figure}[b]
\hbox to \hsize{\epsfxsize=1.0\hsize\hfil\epsfbox{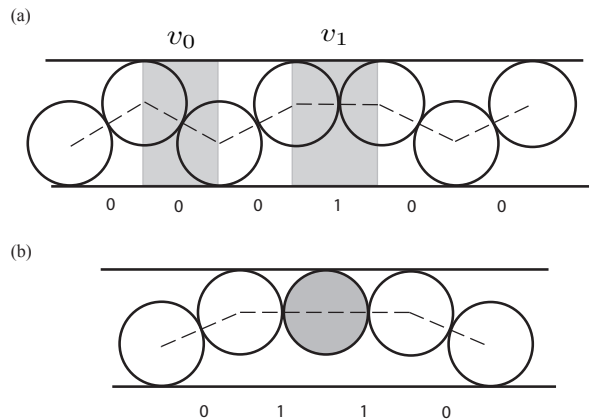}}
\caption{(a) A jammed configuration. The dashed lines  join the centers of contacting discs and are labelled 0, for the most dense, or 1 for the least dense (defect) arrangement. The shaded regions represent the volumes associated with each. (b) An arrangement with two neighboring defects (-1-1-) is not jammed because the shaded disc does not fulfill the local jamming condition and is able to move vertically. }
\label{fig:model}
\end{figure}

Fig.~\ref{fig:model}a shows a jammed configuration containing the two possible local disc arrangements. The most dense state involves two particles contacting across the channel while the {\it defect} state consists of two neighboring particles contacting along the channel, which results in a less dense arrangement. Both the most dense local arrangement and the defect have well defined volumes, $v_0=H\sqrt{(2\sigma-H)H}$ and $v_1=H\sigma$ respectively, that are additive so the volume function, $W$, can be written as,
\begin{eqnarray}
V=W&=&(N-M)v_0+Mv_1\nonumber\\
&=&NH\left[\sqrt{(2-H)H}(1-\theta)+\theta\right]\mbox{,}
\label{eq:v0}
\end{eqnarray}
where $M$ is the number of defects in the system, $\theta=M/N$ is the fraction of defects and $\sigma$ is the unit of length. The occupied volume fraction for a packing is $\phi_J=\pi/(4H[\sqrt{(2-H)H}(1-\theta)+\theta])$. 

%fig 1 model

To count the number of packings with a given $V$, we develop a lattice gas, or Ising model-like, description of each configuration of discs by drawing a bond between the centers of neighboring discs and labelling the defect bonds as ``1" and the most dense bonds as ``0" (see Fig~\ref{fig:model}).  This means the total number of bonds equals $N$. Following Hill~\cite{hill}, if $M$ is the number of defect bonds, we can divide a configuration into blocks consisting of consecutive ones in the chain and blocks consisting of consecutive zeros in the chain. A block of ones is necessarily separated from a block of zero by a boundary consisting of a 1-0 or 0-1 bond. The total number of configurations with $M$ defects and $M_{01}$ boundaries is obtained by considering the number of different ways of arranging $M$ ones among the  $(M_{01}+1)/2$ possible block of ones, such that there is at least one ``1" in a block, and $(N-M)$ zeros among the $(M_{01}+1)/2$ possible block of zeros. This gives, 
\begin{equation}
\Omega_J(M,M_{01})=\frac{M!(N-M)!}     {[M-\frac{M_{01}}{2}]![N-M-\frac{M_{01}}{2}]![\frac{M_{01}}{2}!]^2}\mbox{,}\\
\label{eq:nj1}
\end{equation}
in the limit of large numbers. The volume of a packing of $N$ discs only depends on $\theta=M/N$ and is independent of $M_{01}$. If the defects could be distributed totally randomly throughout the system, then Eq.~\ref{eq:nj1} would reduce to the usual Ising model expression, $\Omega_J(V)=N!/M!(N-M)!$. However, not all defect arrangements result in truly jammed states. When two defects appear next to each other in a ``1-1" arrangement, the central particle does not satisfy the local jamming condition (Fig.~\ref{fig:model}b). This arrangement represents a saddle point in configuration space that will collapse if the central particle is perturbed in a direction normal to the wall. These configurations can be eliminated from the distribution by setting $M_{01}=2M$, which ensures that every defect is isolated from the other defects and gives,
\begin{equation}
\Omega_J(V)=\frac{(N-M)!}{M!(N-2M)!}\mbox{.}\\
\label{eq:nj2}
\end{equation}
The distribution of packings is binomial with a single most dense structure, containing no defects, a single least dense structure with $\theta=0.5$ and a maximum number of packings when $\theta=1/2-\sqrt{5}/10$. 

Mechanical equilibrium is an important property of granular packings that can be understood in terms of a balance between the total degrees of freedom in a system and the number of force equations that constrain or counteract them~\cite{levJR}. In our system, each disc has two translational degrees of freedom, resulting in a total of $2N$ degrees of freedom. Each disc in a jammed packing contacts two other discs and the wall, which results in a total of $N$ disc - disc force equations, noting each contact involves two discs, and $N$ disc - wall force equations, giving a total of $2N$ equations. The number of degrees of freedom is exactly balanced by the number of force equations which shows that all the jammed packings for this model are isostatic, independent of the number of defects. Furthermore, since each particle only contacts one of its neighbors on either side, it does not feel any interaction from particles beyond its nearest neighbours so there is no interaction between defect states. As a result, the properties of the collectively jammed packings, with the same volume, should be independent of how the defects are distributed i.e. there is no distinction between a packing with an ordered arrangement of defects compared to one with a more random arrangement. These properties strongly suggest that the assumption of equal probability for states of equal volume is valid for the current model.

\section{The Second Law of Thermodynamics}

%remarks concerning entropy here.
The second law, in classical thermodynamics, postulates the existence of entropy as a state function in order to understand the driving force behind spontaneous processes. Furthermore, it states that $dS\geq 0$ for any spontaneous process in an {\it isolated} system so that the entropy is maximized when the system is in equilibrium. The most widely used example of the $2^{nd}$ law involves bringing two metal bars at different initial temperatures, say $T_1 < T_2$, into thermal contact. If the composite system is isolated, then experience (experiment) tells us that heat will flow from the hot bar into the cold bar until it comes to equilibrium and the temperature in each bar is the same. 

We can construct an analogous experiment with our granular system as follows: First we note, on the basis of Eqs.~\ref{eq:s2} and \ref{eq:nj2}, that the entropy of the system with a given volume is,
\begin{eqnarray}
\frac{S}{N\lambda}&=&(1-\theta)\ln(1-\theta)-(1-2\theta)\ln(1-2\theta)\nonumber\\
&&-\theta\ln\theta\mbox{,}
\label{eq:s0}
\end{eqnarray}
and the compactivity is,
\begin{eqnarray}
X&=&\frac{\partial V}{\partial S}=\frac{\partial V}{\partial \theta}\frac{\partial \theta}{\partial S}\nonumber\\
&=&\frac{H \left(\sqrt{(2-H) H}-1\right)}{\ln (1-\theta )+\ln
   (\theta )-2 \ln (1-2 \theta )}\mbox{,}
\label{eq:x2}
\end{eqnarray}
where we have made use of the derivatives of Eqs. \ref{eq:v0} and \ref{eq:s0} with respect to $\theta$. In the lower limit, $X\rightarrow 0$ as $\theta\rightarrow 0$, i.e. when the system is in its most dense state. In the upper limit, $X\rightarrow\infty$ as $\theta\rightarrow 1/2-\sqrt{5}/10$, which occurs at the maximum in the distribution of jammed packings. Packings with higher concentrations of defects are not sampled in equilibrium because these would give rise to negative compactivities.

To prepare two subsystems of $N_1$ and $N_2$ discs, at compactivities $X_1$ and $X_2$ respectively, we must in principle, place each subsystem in contact with a ``thermal" reservoir with which it can exchange volume until equilibrium is reached. However, since both the volume and compactivity are functions of the fraction of defects, it is sufficient to select starting configurations with the initial defect concentrations $\theta^{\prime}_1$ and $\theta^{\prime}_2$, which correspond to some initial conditions with $X_1\neq X_2$. We can now bring the two subsystems into contact so that they can exchange volume. This can be achieved by placing the last particle of system one and the first particle of system two in contact so that the configuration remains jammed. In the context of the present experiment, an isolated system is one that cannot exchange volume with an external reservoir so the composite system has a fixed total volume $V_T=V_1+V_2$, where we use large enough system sizes that the small volumes associated with the end effects can be ignored. 
 
The challenge of dealing with a realistic set of dynamics, such as shaking, will be address in more detail in our discussion, but in the meantime we need to use an idealized dynamics that will allow the system to jump from jammed state to jammed state under the conditions of fixed volume.  As the system is perturbed, we expect it to move between jammed states by a series of random local particle rearrangements which will require the movement of defects. If all the particles are made from the same material, our physical intuition tells us that the defects will {\it spontaneously} move from a region of high concentration to a region of low concentration, until eventually the system comes to equilibrium and the defects are, on average, equally distributed throughout. Since both subsystems are made from the same material and have the same size, a uniform distribution of defects also corresponds to a uniform compactivity.

Application of the second law of thermodynamics implies that the equilibrium distribution of defects can be obtained by maximizing the total entropy of the composite system, $S_T=S_1(\theta_1)+S_2(\theta_2)$. The conservation of volume gives 
\begin{equation}
\theta_2=\frac{N_1(\theta^{\prime}_1-\theta_1)+N_2 \theta^{\prime}_2}{N_2}\mbox{,}
\label{eq:t2}
\end{equation}
which can be used in Eq. \ref{eq:s0} to yield an expression for $S_T$ solely in terms of $\theta_1$ and the initial defect concentrations consistent with the starting compactivities for each subsystem. The equilibrium value of $\theta_1$ in the composite system, obtained from $dS_T/d\theta_1=0$, is then
\begin{equation}
\theta_1=\frac{N_1\theta^{\prime}_1+N_2\theta^{\prime}_2}{N_1+N_2}\mbox{.}\\
\label{eq:t1eq}
\end{equation}
This is just the average number of defects, or uniform distribution, which is consistent with our physical expectation. Furthermore, the compactivity in both subsystems are the same at equilibrium. 

\begin{figure}
\hbox to \hsize{\epsfxsize=1.0\hsize\hfil\epsfbox{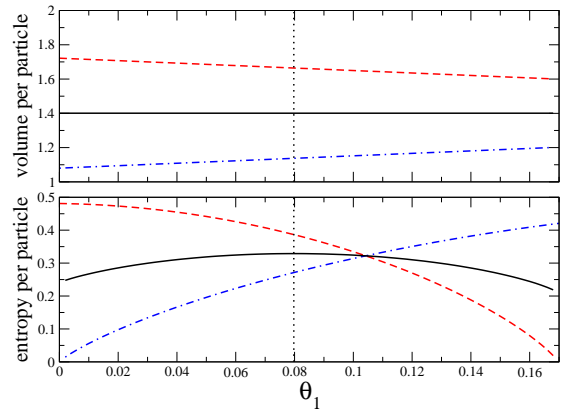}}
\caption{(top) The volume per particle for each subsystem, $V_1/N_1$ (dash-dot line), $V_2/N_2$ (dashed line) and the composite system, $V_T/N$ (dash-dot line). (bottom) The entropy per particle,  $S_1/N_1$ (dash-dot line) $S_2/N_2$ (dashed line) and the composite system, $S_T/N$ (dash-dot line). The vertical dotted line locates the maximum in $S_T$, $N_1=N_2=100$, $\sigma_1=1$, $\sigma_2=1.1\sigma_1$, $H=1.8\sigma_1$, $\theta^{\prime}_1=0$ and $\theta^{\prime}_2=0.27$} 
\label{fig:mix}
\end{figure}

% mixture
The present model can also be used to study the effects of bringing different types of granular materials into contact. The simplest example to study is the case where a subsystem of $N_1$ discs with diameter $\sigma_1$ is brought into contact with a second subsystem of $N_2$ discs with diameter $\sigma_2$, under the conditions $\sigma_1 < \sigma_2$ and $H/\sigma_1< 1+\sqrt{3/4}$, which ensures the interface between the two subsystems can still be made through a single nearest neighbor contact. $V_1$ is given by Eq.~\ref{eq:v0} with $N=N_1$ and the volume of subsection 2 is,
\begin{equation}
V_2=N_2H\left[\sqrt{(2\sigma_2-H)H}(1-\theta)+\sigma_2\theta\right]\mbox{.}
\label{eq:v2m}
\end{equation}
At a fixed $V_T$, we now have,
\begin{equation}
\theta_2=\theta^{\prime}_2-\frac{N_1\left(\theta_1-\theta^{\prime}_1\right)\left(\sqrt{(2-H)H}-1\right)}{N_2\left(\sqrt{(2\sigma_2-H)H}-\sigma_2\right)}\mbox{ ,}\\
\label{eq:t2mix}
\end{equation}
which reduces to Eq.~\ref{eq:t2} as $\sigma_2\rightarrow \sigma_1=1$ and can be used in Eq.~\ref{eq:s0} to provide an expression for $S_2$ in terms of $\theta_1$. Fig.~\ref{fig:mix} shows the volume and the entropy per particle for the two subsystems and the composite system with the initial conditions chosen so that $\theta^{\prime}_1=0$ and $\theta^{\prime}_2=0.27$ which corresponds to $X_1=0$ and $X_2=\infty$, respectively. At equilibrium, where $S_T$ is at a maximum, $X_1=X_2=0.318$, but neither the volumes or entropies of the two subsystem are equal because the change in volume associated a defect is different in each subsystem as a result of the different particle sizes.  
% mixture fig.

To show that the compactivities in the two subsystems are equal when $S_T$ is maximized at fixed $V_T$, we begin by assuming $X_1\neq X_2$ and show that this leads a contradiction to the conservation of the total volume. From Eq.~\ref{eq:x2} we have,
\begin{equation}
X_1=\left(\frac{\partial V_1}{\partial \theta_1}\right)\left(\frac{\partial \theta_1}{\partial S_1}\right)\neq \left(\frac{\partial V_2}{\partial \theta_2}\right)\left(\frac{\partial \theta_2}{\partial \theta_1}\right)\left(\frac{\partial \theta_1}{\partial S_2}\right)=X_2
\label{eq:x1x2}\mbox{ ,}
\end{equation}
where we have used the fact that $\theta_2$ is a function of $\theta_1$ at fixed $V_T$. Maximizing the total entropy yields
\begin{equation}
\left(\frac{\partial S_1}{\partial \theta_1}\right) = -\left(\frac{\partial S_2}{\partial \theta_1}\right)\mbox{ ,}\\
\label{eq:s1s2}
\end{equation}
which, in combination with Eq.~\ref{eq:x1x2} gives,
\begin{equation}
\left(\frac{\partial V_1}{\partial \theta_1}\right )\neq -\left(\frac{\partial V_2}{\partial \theta_2}\right)\left(\frac{\partial \theta_2}{\partial \theta_1}\right)=-\left(\frac{\partial V_2}{\partial \theta_1}\right)\mbox{ .}\\
\label{eq:proof}
\end{equation}
This completes the proof as this last inequality suggests that the volume lost from one system is not equal to the volume gained by the other as the number of defects in system 1 is varied which violates the conservation of volume. Therefore $X_1$ must equal $X_2$ at equilibrium.

\section{Canonical Equilibrium}
The equivalent to the canonical ensemble in a thermal system is obtained by integrating over all possible degrees of freedom and providing a Boltzmann type weight to each jammed state on the basis of its volume, which for the present system gives,
\begin{eqnarray}
e^{\frac{-Y}{\lambda X}}&=&\int \Theta e^{-W/\lambda X} d\mbox{(all degrees of freedom)}\nonumber\\
&=&\sum_{M=0}^{M=N/2}\Omega(M,N;V)e^{-W/\lambda X}\nonumber\\
&=& e^{\frac{-v_0N}{\lambda X}} \, _2F_1\left[\frac{1}{2}-\frac{N}{2},-\frac{N}{2};-N;-4
   e^{\frac{v_0}{\lambda X}-\frac{H}{\lambda X}}\right]\mbox{,}
\label{eq:can}
\end{eqnarray}
where $Y$ is the analog to the free energy, $_2F_1$ is a hypergenometric function and the integral has been reduced to a sum over the discrete set of jammed states. The free energy can now be used in conjunction with the equivalent thermodynamic machinery to obtain other thermodynamic variables. For example, in analogy with the Helmholtz free energy, given
\begin{equation}
Y=V-XS\mbox{,}\\
\label{eq:y}
\end{equation}
the entropy of the system can obtained from
\begin{equation}
S=-\left(\frac{\partial Y}{\partial X}\right)_V\mbox{ .}
\label{eq:ts}
\end{equation}

%fig 2 Entropy and volume as function of X
\begin{figure}
\hbox to \hsize{\epsfxsize=1.0\hsize\hfil\epsfbox{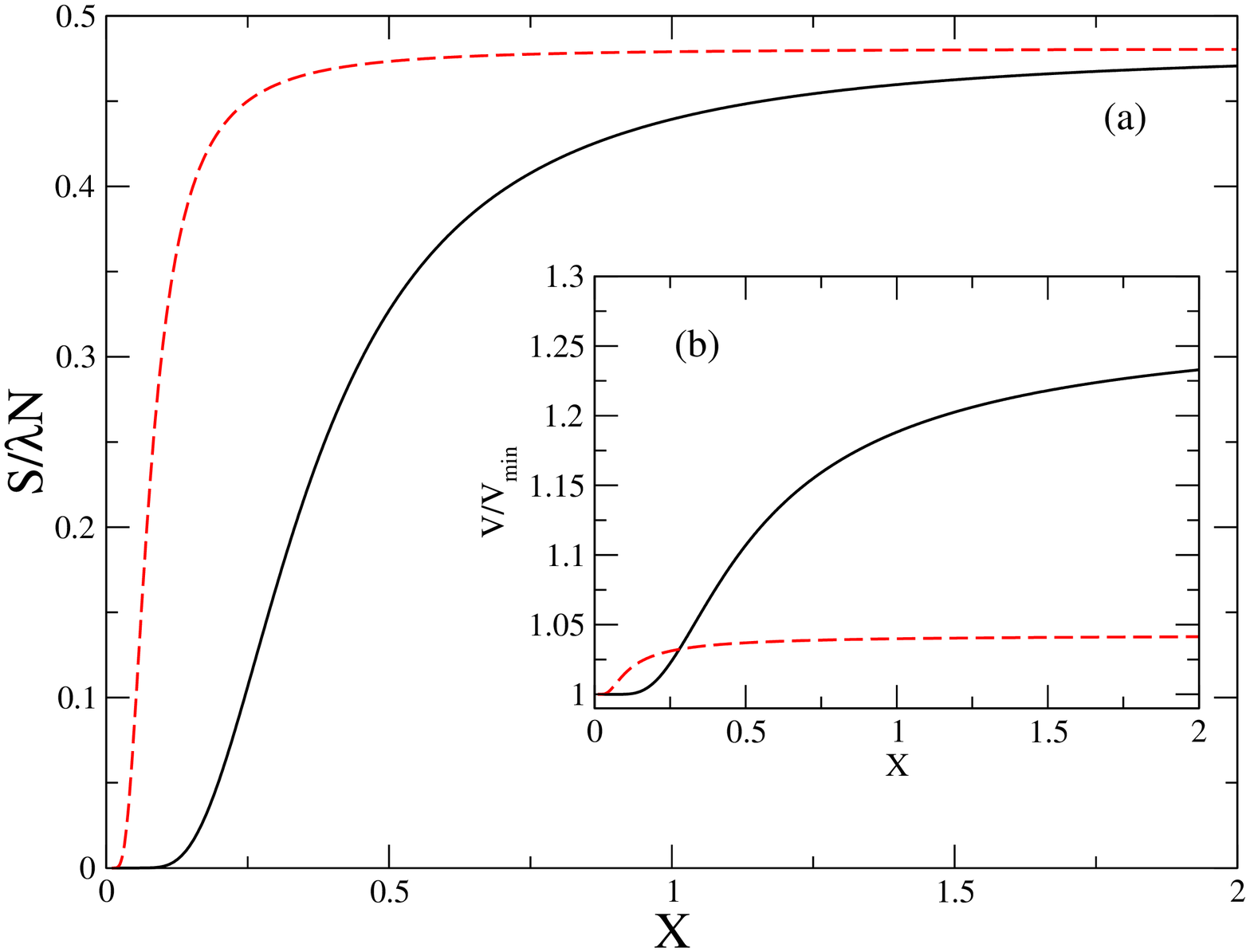}}
\caption{(a) The entropy, $S$, and (b) the volume relative to the volume of the most dense state, $V/V_{min}$, as a function $X$ for a system with particles $H/\sigma=1.86$ (solid line) and $H/\sigma=1.50$ (dashed line).} 
\label{fig:2}
\end{figure}
Fig.~\ref{fig:2} shows the entropy and volume relative to the most dense packing for a granular system of discs confined within channels of diameter $H/\sigma=1.86$ and $H/\sigma=1.50$ as a function of the compactivity. The underlying distributions of jammed states is the same for both systems and the difference in behavior between the two arises from the larger change in volume associated with the gain or loss of a defect state in the wider channel. As a result, the relative change in volume is greater, and occurs  gradually as a function of $X$, for the wider channel. The properties of the narrow channel are relatively insensitive to changes in $X$ at the higher compactivities, but we see a rapid decrease in entropy and volume as $X$ approaches zero.

\section{Discussion}
The goal of the present paper was to explore the statistical mechanics of a simple model granular system. We are able to carry out a complete analysis of the distribution of jammed states for a system of highly confined hard discs and find that all the packings of a fixed volume are structurally the same. This suggests that the Edwards' assumption of equi-probability for jammed states of equal volume is well justified in the present model, even if it is not valid for more complex granular systems. The model also represents a genuine granular system that can be realized experimentally and it is rare to find cases where the exact results for real experiments are available.

However, there are aspects of granular systems not covered in this model. First, we point out that our analysis required an idealistic dynamics that allowed the system to move directly between jammed states at fixed volume. Shaken granular materials need to generate a momentary increase in free volume to allow the particles to unjam, move and then return to a jammed state. 
The role of the energy injection and how the free volume affects the way the set of jammed states is sample is a critical feature of granular systems that has not been included here. The presence of friction between particles has also been ignored and we would expect this to alter the number and type of packings included in the distribution, as well as possibly influencing the assumption of equi-probable volume states. For example, it is immediately obvious that the neigboring defect state (Fig.~\ref{fig:model}b) would become stable in the presence of friction, but other   ``bridge-type" particle contacts may also need to be included. Nevertheless, the simplicity of the present system and its accessibilty to experiment, suggest that it may be an invaluable tool in understanding some of the elementary aspects in the statistical mechanics of granular systems.

\end{document}